\def\ube13{UBe$\rm_{13}$}

\def\bi2212{Bi$\rm_2$Sr$\rm_2$CaCu$\rm_2$O$\rm_8$}
\def\ybi2212{Bi$\rm_2$Sr$\rm_2$YCu$\rm_2$O$\rm_8$}
\def\ycabi2212{Bi$\rm_2$Sr$\rm_2$Ca$\rm_{1-x}$Y$\rm_x$Cu$\rm_2$O$\rm_{8+\delta}$}

\def\y65cabi2212{Bi$\rm_2$Sr$\rm_2$Ca$\rm_{0.35}$Y$\rm_{0.65}$Cu$\rm_2$O$\rm_{8+\delta}$}
\def\Ir{CeIrIn$_5$}
\def\Co{CeCoIn$_5$}

\documentstyle[aps,prl,twocolumn,epsf]{revtex}
\begin{document} 
\draft

\def\dfrac#1#2{{\displaystyle{#1\over#2}}}
\twocolumn[\hsize\textwidth\columnwidth\hsize\csname @twocolumnfalse\endcsname
Submitted to PRL \hfill{LA-UR-00-5563}

\title{Unconventional superconductivity in CeIrIn$_5$ and CeCoIn$_5$: Specific heat and thermal conductivity studies}

\author{R. Movshovich,$^1$ M. Jaime,$^1$ J. D. Thompson,$^1$ C. Petrovic,$^2$ Z. Fisk,$^2$ P. G. Pagliuso,$^1$ and J. L. Sarrao$^1$ }
\address{$^1$Los Alamos National Laboratory, Los Alamos, New Mexico 87545 \\ $^2$NHMFL, Florida State University, Tallahassee, FL 32306}

\date{\today}

\maketitle

\begin{abstract} 

Low temperature specific heat and thermal conductivity measurements on the ambient pressure heavy fermion superconductors \Ir\  and \Co\  reveal power law temperature dependences of these quantities below T$_c$. The low temperature specific heat in both \Ir\ and \Co\ includes T$^2$ terms, consistent with the presence of nodes in the superconducting energy gap. The thermal conductivity data present a T-linear term consistent with the universal limit (\Ir), and a low temperature T$^3$ variation in the clean limit (\Co), also in accord with prediction for an unconventional superconductor with lines of nodes. 

\end{abstract}


 ]
\narrowtext

Unconventional superconductivity has been an active area of research for several decades, ever since the discovery of the first heavy fermion superconductor, CeCu$_2$Si$_2$~\cite{steglich79:cecu2si2}. The presence of strong magnetic interaction between 4f moments and itinerant electrons in this class of compounds allows the possibility of non-phonon mediated coupling between superconducting quasiparticles, a signature of unconventional superconductivity, and a superconducting order parameter with lower symmetry than that of the underlying crystal lattice. Soon after the discovery of superconductivity in CeCu$_2$Si$_2$, several uranium-based heavy fermion superconductors were discovered. The presence of a double transition in UPt$_3$ immediately identified this compound as an unconventional superconductor~\cite{fisher:prl_89}. Subsequent observations of power law temperature dependences of the specific heat and thermal conductivity have been instrumental in identifying UPt$_3$ (and other U-based heavy fermion superconductors) as unconventional.

Until very recently, among Ce-based heavy fermion compounds only CeCu$_2$Si$_2$ was shown to superconduct at ambient pressure.  Other ThCr$_2$Si$_2$-based compounds require application of significant pressure (on the order of 20 kbars) before they exhibit superconductivity. These include CeCu$_2$Ge$_2$~\cite{jaccard92:CeCu2Ge2}, CePd$_2$Si$_2$~\cite{grosche:physB_96,mathur:nature_98}, and CeRh$_2$Si$_2$~\cite{movshovich:prb_96}. Recently, cubic CeIn$_3$ was also shown to superconduct under pressure of about 25 kbar~\cite{walker:physicaC_97}, with superconductivity mediated by magnetic interactions~\cite{mathur:nature_98}. 

Quantitative thermodynamic and heat transport measurements, that probe the nature of the superconducting gap, are a lot simpler at ambient pressure. CeCu$_2$Si$_2$ seems to be a good candidate for such investigations. However, it is a rather complicated system, in which very small changes in stoichiometry can change the ground state from superconducting to antiferromagnetic. Recently discovered ambient pressure superconducting \Ir~\cite{petrovic:epl_01} and \Co~\cite{petrovic:nature_00} do not display such complications. Therefore, they present an uncommon opportunity to study the superconducting ground state of Ce-based heavy fermion compounds.  In this Letter we present the results of low temperature specific heat and thermal conductivity measurements on both \Ir\ and \Co\, which show power law temperature dependences and present a strong proof of the unconventional nature of the superconductivity in these compounds.

The details of sample growth and characterization are described in Refs.~\onlinecite{petrovic:epl_01,petrovic:nature_00}. Large  plate-like single crystals, up to 1 cm long, are grown from an excess In flux. Their quasi-2D tetragonal structure can be viewed as layers of CeIn$_3$ separated by layers of IrIn$_2$ (or CoIn$_2$). Therefore, we can treat CeIn$_3$ as the parent compound for \Ir\  and \Co. Both \Ir\ and \Co\ 
are ambient pressure superconductors, with transition temperature 

\begin{figure}
\epsfxsize=3in
\centerline{\epsfbox{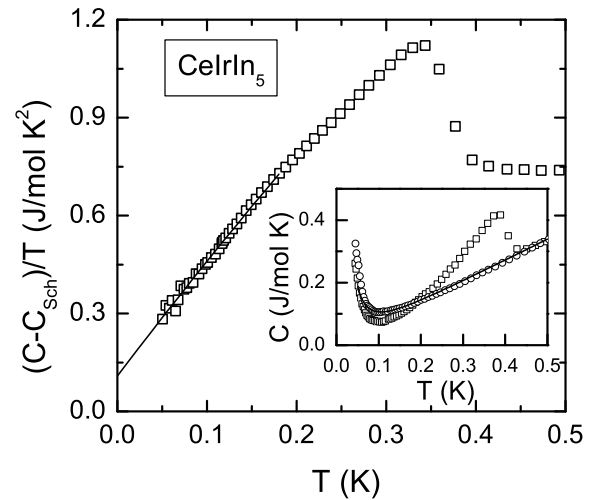}}
\caption{Specific heat of \Ir, as $(C-C_{Sch})/T$ vs. T. ($\Box$) H = 0 kG, ($\circ$) H = 5 kG. Dashed line is $\gamma = C/T = 0.68$ $\rm J/mol K^2$; solid line is a fit to H = 0 kG data for  85 mK $< T <$ 0.2 K. Inset: $C(T)$, ($\Box$) H = 0 kG, ($\circ$) H = 5 kG. Solid line is a sum of $\gamma T$ and nuclear Schottky $C_{Sch}$ for H = 5 kG data. } 
\label{Ir_hc}
\end{figure}

\begin{figure}
\epsfxsize=3in
\centerline{\epsfbox{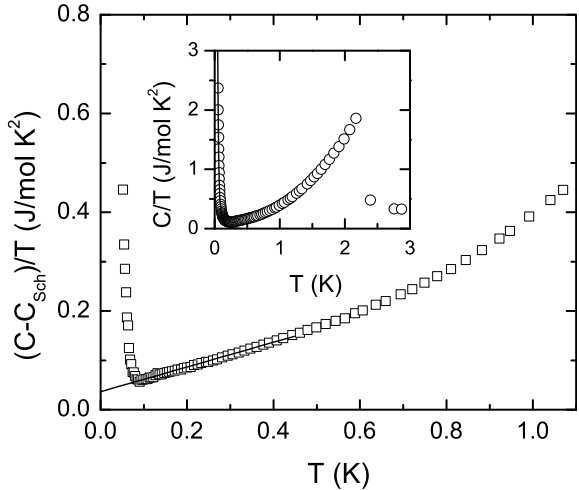}}
\caption{$(C - C_{Sch})/T$ of \Co\ at H = 0 kG. Solid line is a fit for 0.1 K $< T <$ 0.7 K. Inset: Specific heat of \Co: Solid line is In nuclear quadrupolar Schottky anomaly $C_{Sch}$.}
\label{Co_hc}
\end{figure}
 
\noindent T$_c$ of 400 mK and 2.3 K and specific heat jump at superconducting phase transition $\Delta C/\gamma T_c$ of 0.76 and 4.3 respectively.
 
Fig.~\ref{Ir_hc} shows specific heat of two samples of \Ir\ in zero field and in magnetic field of 5 kG (${\bf H}\parallel {\bf c}$), which suppresses the superconducting state to $T = 0$ K. The solid line in the inset represents the sum of the T-linear term expected for a well developed heavy fermion state and the In nuclear quadrupolar Schottky anomaly $C_{Sch}$, and it agrees well with the 5 kG normal state data for \Ir. The five doubly degenerate energy levels used in calculating $C_{Sch}$ (In nuclear spin $I = 9/2$) were measured directly via NQR experiment~\cite{curro:unpublished_00}. 

The main body of Fig.~\ref{Ir_hc} shows specific heat divided by temperature after $C_{Sch}$ was subtracted.  
The solid line in the main body of Fig.~\ref{Ir_hc} is a linear fit to the data in the superconducting state of \Ir\ below 0.2 K ($T_c/2$). The vertical intercept gives the coefficient of the T-linear term $\gamma_0 = .110 \pm .010$ $\rm J/mol K^2$ (a second sample shown in the inset gives $\gamma_0 = .010 \pm .008$ $\rm J/mol K^2$ between 200 mK and 85 mK). 
Low temperature T-linear contribution in specific heat of unconventional superconductors is commonly attributed to the impurity band that forms in the linear node(s) of the superconducting energy gap. The sample with larger $\gamma_0$ has lower T$_c$ of 0.38 K, compared to 0.4 K for the sample with lower $\gamma_0$, qualitatively consistent with the impurity band origin of the linear-T term in specific heat. The $T^2$ term in electronic specific heat in \Ir\ is an indication of the presence of lines of nodes in the energy gap, as in a number of other unconventional superconductors, including both cuprates YBCO~\cite{moler:prl_94} and LSCO~\cite{momono:physC_96} and heavy fermion UPd$_2$Al$_3$~\cite{sakon:physicaB_94}. The coefficient of the T-squared term in the specific heat, $\alpha = 3.56 \pm 0.06$ $\rm J/mol K^3$, of \Ir\ is several orders of magnitude greater than that of the HTS, reflecting the heavy-fermion nature of this compound. 

Specific heat of \Co\ in zero field is shown in Fig.~\ref{Co_hc}. The inset of Fig.~\ref{Co_hc} shows the data up to 3 K together with calculations of the In quadrupolar nuclear Schottky anomaly (solid curve) based on the energy levels determined via NQR studies of \Co~\cite{Simovic:unpublished_00}. The good agreement between calculation and data leaves little doubt that the low temperature anomaly is due to In nuclear quadrupolar moments. In the main body of Fig.~\ref{Co_hc} we show specific heat divided by temperature after subtracting the In nuclear Schottky contribution. The zero field data are well described by $C/T = 0.04$ ${\rm J/mol K^2} + 0.25 T$ $\rm J/mol K^3$ for 95 mK $ < T <400$ mK suggesting the presence of impurity band and line(s) of nodes in the energy gap, as in the case of \Ir. The low temperature upturn in the data may be related to the low temperature entropy revealed when 5 T field is applied to suppress superconductivity in this compound.~\cite{petrovic:nature_00} Crystallographically, \Ir\ and \Co\ are closely related compounds. The similar temperature dependences of their specific heats points (gratifyingly) to the  possibility of {\it the same} symmetry of their superconducting order parameters. 

\begin{figure}
\epsfxsize=3in
\centerline{\epsfbox{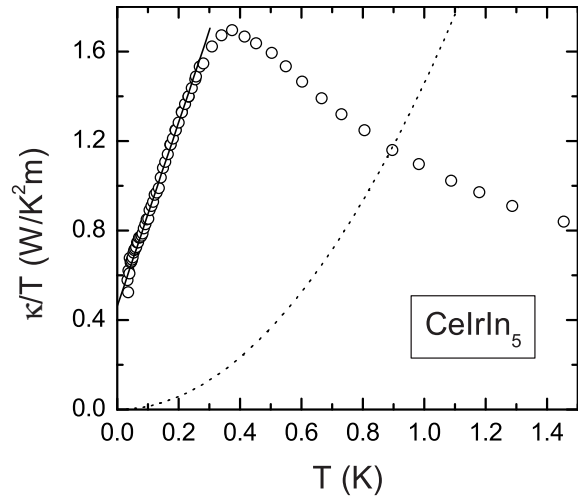}}
\caption{Thermal conductivity of \Ir. Solid line is a linear fit to the data for $T < 0.2$ K = T$_c$/2. Dotted line is an upper limit estimate for the phonon thermal conductivity.}
\label{Ir_tc}
\end{figure}

We now turn attention to thermal transport in \Ir\ and \Co. Thermal conductivity $\kappa$ of \Ir\ divided by temperature is displayed in Fig.~\ref{Ir_tc} for 2 K $< T < 33$ mK. $\kappa / T$  reaches a maximum at $T_c = 0.4$ K, begins to fall below $T_c$, and becomes linear in temperature below $T_c / 2 = 200$ mK. The solid line is a fit $\kappa/T = .46 \pm .07$ $\rm W/K^2m$ $+ (4.10 \pm .06)\rm T$ ${\rm W/K^3m}$ for $T < 200$ mK. The 
dashed line in Fig.~\ref{Ir_tc} is an upper limit estimate of the phonon thermal conductivity (with boundary scattering only) $\kappa = {1\over 3} \beta T^3 \langle v \rangle \Lambda_0$. 
Here $\beta T^3 = 15.5  T^3$ $\rm {J \over {m^3 K^4}}$ is the phonon specific heat of the non-magnetic analogue LaIrIn$_5$~\cite{hundley:unpublished_00}, $\Lambda_0$ is the phonon mean free path, and $ \langle v \rangle$ is the mean phonon velocity~\cite{thacher:pr_67,phonon_tc_limit}. On the basis of this 

\noindent estimate we can conclude that the heat transport below T$_c$ is dominated by quasiparticles.

The presence of a T-linear term in $\kappa$ is expected for an unconventional superconductor with line(s) of nodes in the energy gap on the Fermi surface. It arises due to the appearance of the impurity band at these nodes. At first sight the magnitude (0.46 $\rm W/K^2m$) of the coefficient of the linear term  seems to be inconsistent with the results of the specific heat measurements, which indicate a very clean sample with low impurity concentration. However, Graf {\it et al.}~\cite{graf:prb_96} has shown that  the low temperature limit of thermal conductivity in unconventional superconductors with a variety of  order parameter symmetries that have lines of nodes is universal (independent of the concentration of scattering impurities), 

\begin{equation}
\left. {\kappa \over T}\right|_{T \rightarrow 0}= {\pi^2 \over 3} N_f v_f^2 \times {a \over {2 \mu \Delta_0}} . \label{eq:universal tc limit}
\end{equation}

Here $N_f$ is the normal density of states, $v_f$ is a Fermi velocity, $\Delta_0$ is a maximum of the energy gap at $T = 0$, $\mu$ is the slope parameter which describes the rate of the increase of the energy gap away from the node on the Fermi surface, and $a$ is 1 for 3D order parameter with lines of nodes and $4 \over \pi$ for the 2D case (as in the d-wave state of the high-temperature superconductors (HTS)). In the rest of this Letter and in variety of calculations of Ref.~\onlinecite{graf:prb_96} $\mu$ was taken to be equal to 2.

We obtain the normal state density of states $N_f$, Fermi velocity $v_f$, and mean free path $l_{tr}$ (or scattering life time $\tau$) at T$_c$ following the BCS-based analysis of Refs.~\onlinecite{orlando:pr_79,details_to_follow}. The input parameters for this analysis are the linear coefficient of the specific heat $\gamma$, the derivative of the critical magnetic field $H_{c2}^\prime$ at T$_c$, and the normal state resistivity at T$_c$ $\rho_0$~\cite{ceirin5_parameters}. We obtain $N_f = 1.13\times 10^{36}$ $\rm (erg cm^3)^{-1}$, $v_f = 7.65\times 10^3$ $\rm m/s$, effective mass $m_{eff} \approx 140 m_e$, and $l_{tr} = 1350$ \AA\ at T$_c$. These  numbers are to be compared with those obtained from dHvA experiments~\cite{haga:unpublished_00}, which have observed $m_{eff}$ of up to $45 m_e$, Fermi velocities down to $6.3 \times 10^3 $ $\rm m \over s$, and quasiparticle mean free paths between 58 \AA\ (for the lightest band) and 4500 \AA\ (for the heaviest band observed). Overall, the results of the analysis above compare favorably to those of dHvA experiments. We also obtain zero temperature coherence length $\xi_0 = \sqrt{\phi_0 \over 2 \pi H_{c2}(T=0)} = 241$ \AA\  from the H-T phase diagram~\cite{petrovic:epl_01}, and the ratio $l_{tr}/\xi_0 = 5.43$, which shows that \Ir\ is indeed in the clean limit. The energy gap is then $\Delta_0/k_B = 0.746$ K, and $\Delta_0 /k_B T_c = 1.865$, close to BCS value of 1.763.

We can now estimate the universal limit via Eq.~\ref{eq:universal tc limit}, obtaining $\left. {\kappa \over T}\right|_{T \rightarrow 0} = 1.06$ $\rm W/K^2m$ for a 3D order parameter. If in addition we take $\Delta_0= 2.14 k_B T_c $, a weak coupling value for the d-wave order parameter in the clean limit, we obtain $\left. {\kappa \over T} \right|_{T\rightarrow 0} = 1.2$ $\rm W/K^2m$ for a 2D d-wave order parameter. Given the fact that the Orlando analysis~\cite{orlando:pr_79} assumes a spherical Fermi surface, while LDA band structure calculations~\cite{wills:unpublished_00} reveal a very complicated Fermi surface with contributions from three different bands, an uncertainty in estimated $\left. {\kappa \over T} \right|_{T\rightarrow 0}$  of about a factor of two seems reasonable. Therefore the large observed $\left. {\kappa \over T} \right|_{T\rightarrow 0}= 0.46$ $\rm W/K^2m$ can be explained on the basis of the theory of Graf {\it et al.}~\cite{graf:prb_96}, and is additional strong proof of the presence of line(s) of nodes in the gap of \Ir. 

\begin{figure}
\epsfxsize=3in
\centerline{\epsfbox{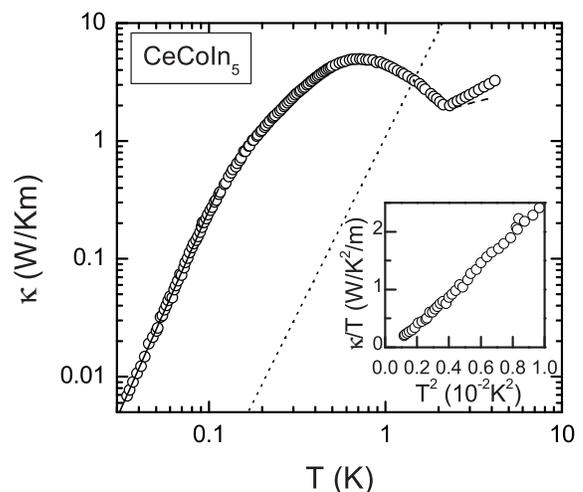}}
\caption{Thermal conductivity of \Co. Solid line is a power law fit for $T < 100$ mK $< T_c/20$. Dotted line is an upper limit estimate of $\kappa_{ph}$. Dashed line is $\kappa_e$ for $T > T_c$. Inset: $\kappa /T$ vs. $T^2$ for $T < 0.1$ K.}
\label{Co_tc}
\end{figure}

Finally, Fig.~\ref{Co_tc} shows thermal conductivity of \Co\ on a log-log plot. The data are striking in several ways. First, upon the sample entering the superconducting state, thermal conductivity displays a sharp kink and rises from 2 $\rm W/Km$ at T$_c = 2.3$ K to a maximum value of 5 $\rm W/Km$ at 0.71 K ($\rm T / T_c = 0.3$), an increase of a factor of 2.5. It is more instructive to consider the value of $\kappa /T$, which is proportional to the product of the density of normal quasiparticles and their life time. $\kappa /T$ grows from 0.86 $\rm W/K^2m$ at T$_c = 2.3 $ K and reaches the value 8.7 $\rm W/K^2m$ at $\rm T = 0.46$ K (T/T$_c$ = 0.2). In spite of the strong reduction in the number of normal quasiparticles expected at $\rm T = 0.2 T_c$, $\kappa /T$ actually grows by more than an order of magnitude! An increase of $\kappa /T$ was also observed in several HTS, where the separation of electronic and phonon contribution, and therefore identification of the origin of the peak in $\kappa$, is rather complicated~\cite{cohn:prb_92,yu:prl_92}. 

In the case of \Co\, the situation is much more straightforward. The dotted line
in Fig.~\ref{Co_tc} is an upper 
limit estimate of the phonon thermal conductivity $\kappa_{ph}$ in \Co. At the temperature of the maximum in $\kappa$ the phonon upper bound is a factor of 15 below the measured value. The dashed line for T $>$ T$_c$ is a Widemann-Franz estimate of electronic thermal conductivity in the normal state $\kappa_e^n = L_0 T / \rho$, with the normal state resistivity $\rho$ from Ref.~\onlinecite{petrovic:nature_00} and the Lawrence number $L_0 = 2.44 \times 10^{-8}$ $\rm W\Omega K^{-2}$. In the normal state the thermal conductivity 

\noindent of \Co\ is dominated by electrons, which contribute over 90\% of the experimental $\kappa$ at T$_c$. We can thus conclude that \textit{all} of the peak below T$_c$ in \Co\ is due to electrons, and may be viewed alternatively as supporting the mostly electronic origin of the peak in $\kappa$ in HTS as well. This peak may bear a direct relation to the appearance of the low temperature rise in the normal state specific heat at $H = 5$ T, which compensates the entropy from a very large specific heat jump ${\Delta C\over C_n} = 4.3$ at T$_c$ in zero field~\cite{petrovic:nature_00}. This large $\Delta C\over C_n$ suggests that the fluctuations in the normal state, responsible for the low temperature tail in $C$ in the 5 T data, are redistributed up to T$_c$ when \Co\ enters the superconducting state. Such fluctuations also scatter electrons, and their suppression in the superconducting state would increase $\kappa / T$.

We estimated the universal limit of $\kappa / \rm T$ in \Co\ via an analysis similar to that for \Ir~\cite{cecoin5_parameters}. The dirty limit term contributes only 6\% to $H_{c2}^\prime$, and ${l_{tr} / \xi_0} \approx 14$ at T$_c$, clearly indicating that \Co\ is also a clean superconductor~\cite{cecoin5_orlando}. Using the values of $\Delta_0$ and $a = 1$ (for a 3D superconductor with lines of nodes) in Eq.~\ref{eq:universal tc limit}, we obtain for \Co\ $\left. \kappa / T \right|_{T\rightarrow 0} = 0.19$ $\rm W/K^2m$. With the d-wave energy gap $\Delta_0 / k_B = 2.14 \rm T_c$ and $a = 4/\pi$ for a 2D superconductor, we obtain $\left. \kappa / T \right|_{T\rightarrow 0} \approx 0.1$ $\rm W/K^2m$. The inset of Fig.~\ref{Co_tc} shows $\kappa / \rm T$ vs. $T^2$  below 100 mK. The minimum value measured is 0.2 $\rm W/K^2m$, which suggests that the sample is close to universal limit. Future measurements to investigate universal limit of \Co\ are planned.

For temperatures below 100 mK (T/T$_c \approx .043$) and down to the lowest temperature measured of 33 mK ($\approx$ 1.5\% of T$_c$) thermal conductivity follows a power law behavior. The straight line in Fig.~\ref{Co_tc} is a fit $\kappa = a T^{3.37}$ to the data. This is close to T$^3$ behavior predicted for unconventional superconductor in a clean limit with line(s) of nodes~\cite{graf:prb_96}. Therefore the heat transport in \Co\ between 33 mK and 100 mK can be described as that of an unconventional superconductor in the clean limit, with an impurity band width of less than 30 mK. An upper limit estimate of the impurity concentration that gives rise to such narrow impurity band~\cite{graf:prb_96} is $n_{imp} = 20$ ppm in the unitary scattering limit and $n_{imp} = 20$ ppt in the Born limit. The true impurity concentration is probably between these limits, and closer to the unitary scattering limit~\cite{petrovic:nature_00}.

In summary, both \Ir\ and \Co\ display power law behavior in specific heat and thermal conductivity, indicative of unconventional superconductivity. T-squared terms in specific heat imply existence of lines of nodes in both superconducting energy gaps. Thermal conductivity also supports unconventional order parameters in these compounds, which manifests itself in (1) large $\left. \kappa / T \right|_{T\rightarrow 0} = 0.46 \rm W/K^2m$ in \Ir\, consistent with the universal limit estimates, (2) large increase in electronic thermal conductivity below $T_c = 2.3K$ in \Co\ reminiscent of the behavior of HTS, (3) close to T$^3$ behavior between 33 mK and $\approx 100 mK$ in \Co, predicted for unconventional superconductors in clean limit. 

We thank M. F. Hundley, G. Lonzarich, F. Steglich, A. V. Balatsky, M. Graf, I. Vekhter, and L. Bulaevsky for stimulating discussions. Work at Los Alamos National Laboratory was performed under the auspices of the U.S. Department of Energy. Work at NHMFL is supported by the NSF grant 502459022.


\begin{thebibliography}{10}

\bibitem{steglich79:cecu2si2}
F. Steglich {\it et~al.}, Phys. Rev. Lett. {\bf 43},  1892  (1979).

\bibitem{fisher:prl_89}
R.~A. Fisher {\it et~al.}, Phys. Rev. Lett. {\bf 62},  1411  (1989).

\bibitem{jaccard92:CeCu2Ge2}
D. Jaccard, K. Behnia, and J. Sierro, Phys. Lett. A {\bf 163},  475  (1992).

\bibitem{grosche:physB_96}
F.~M. Grosche, S.~R. J. N.~D. Mathur, and G.~G. Lonzarich, Physica B {\bf 224},
   50  (1996).

\bibitem{mathur:nature_98}
N.~D. Mathur {\it et~al.}, Nature {\bf 394},  39  (1998).

\bibitem{movshovich:prb_96}
R. Movshovich {\it et~al.}, Phys. Rev. B {\bf 53},  8241  (1996).

\bibitem{walker:physicaC_97}
I.~R. Walker, F.~M. Grosche, D.~M. Freye, and G.~G. Lonzarich, Physica C {\bf
  282},  303  (1997).

\bibitem{petrovic:epl_01}
C. Petrovic {\it et~al.}, Europhys. Lett. {\bf 53},  354  (2001).

\bibitem{petrovic:nature_00}
C. Petrovic {\it et~al.}, submitted to Nature  .

\bibitem{curro:unpublished_00}
N. Curro, unpublished  .

\bibitem{moler:prl_94}
K.~A. Moler {\it et~al.}, Phys. Rev. Lett. {\bf 73},  2744  (1994).

\bibitem{momono:physC_96}
N. Momoto and M. Ido, Physica C {\bf 264},  311  (1996).

\bibitem{sakon:physicaB_94}
T. Sakon {\it et~al.}, Physica B {\bf 199},  154  (1994).

\bibitem{Simovic:unpublished_00}
B. Simovic, unpublished  .

\bibitem{hundley:unpublished_00}
M.~F. Hundley, unpublished  .

\bibitem{thacher:pr_67}
P.~D. Thacher, Phis. Rev. {\bf 156},  975  (1967).

\bibitem{phonon_tc_limit}
$\Lambda_0 = {2\over \pi} \sqrt{a b} = 140$ $\rm \mu m$, $a$ and $b$ are sample
  crossection dimensions, $ \langle v \rangle = v_l {{2s^2+1} \over {
  2s^3+1}}$, where $s = v_l/v_t$ is the ratio of the longitudinal and
  transverse velocities. Bulk modulus (assumed here to be $B = 800$ kbar)
  yields bulk velocity $v_{bulk} = {\sqrt {B /\rho}} \approx 3000$ m/s.
  Available data on tetragonal CeCu$_2$Si$_2$ (scaled by the bulk modulus),
  gives $1500 < \langle v \rangle < 2000$ m/s. We assume $\langle v \rangle =
  2000$ m/s.

\bibitem{graf:prb_96}
M.~J. Graf, S.-K. Yio, J.~A. Sauls, and D. Rainer, Phys. Rev. B {\bf 53},
  15147  (1996).

\bibitem{orlando:pr_79}
T.~P. Orlando {\it et~al.}, Phys. Rev. {\bf 19},  4545  (1979).

\bibitem{details_to_follow}
details of this analysis and its additional results will be published
  elsewhere.

\bibitem{ceirin5_parameters}
For CeIrIn$5$, $\gamma = 0.7$ $\rm J/mol K^2$ $= 7.1\times 10^3$ $\rm J/m^3
  K^2$, $H_{c2}^\prime = 2.54$ T/K, and $\rho_0 = 1$ $\rm \mu \Omega - cm$.

\bibitem{haga:unpublished_00}
Y. Haga {\it et~al.}, unpublished  .

\bibitem{wills:unpublished_00}
J. Wills, unpublished  .

\bibitem{cohn:prb_92}
J.~L. Cohn {\it et~al.}, Phys. Rev. B {\bf 45},  13144  (1992).

\bibitem{yu:prl_92}
R.~C. Yu, M.~B. Salamon, J.~P. Lu, and W.~C. Lee, Phys. Rev. Lett. {\bf 69},
  1431  (1992).

\bibitem{cecoin5_parameters}
For CeCoIn$_5$, $\gamma = 0.29$ $\rm J/mol K^2$ $= 3\times 10^{3}$ $\rm J /
  m^3K^2$, $H_{c2}^\prime = 8.18$ T/K, $\rho_0 = 3.1$ $\rm \mu \Omega - cm$.

\bibitem{cecoin5_orlando}
We also obtain $N_f = 4.774\times 10^{35}$ $\rm (erg cm^3)^{-1}$, $v_f =
  9.74\times 10^3$ $\rm m/s$, $m_{eff} \approx 83 m_e$, $l_{tr} = 810$ \AA\ at
  T$_c$, $\xi_0 = 58 \AA$, $\Delta_0 = 4.03\times 10^{-16}$ erg, and ${\Delta_0
  \over k_B T_c} = 1.27$.

\end{thebibliography}

\end{document}